\documentclass[review]{elsarticle}

\usepackage{lineno,hyperref}
\modulolinenumbers[5]
\usepackage{multirow}
\usepackage[utf8]{inputenc}

\journal{Journal of \LaTeX\ Templates}

\newcommand{\invnb}{{nb}$^{-1}$}
\newcommand{\gev}{GeV/$c$}
\newcommand{\gevtwo}{GeV/$c^{2}$}

\newcommand{\jpsi}{\ensuremath{J/\psi}}

\newcommand{\AuAu}{Au+Au}
\newcommand{\PbPb}{Pb+Pb}
\newcommand{\pp}{$p$+$p$}

\newcommand{\sqrtsNN}{\ensuremath{\sqrt{s_{\mathrm {NN}}}}}
\newcommand{\sqrts}{\ensuremath{\sqrt{s}}}

\newcommand{\pT}{\ensuremath{p_\mathrm{T}}}

\newcommand{\nsigmapi}{\ensuremath{\mathrm{n}\sigma_{\pi}}}
\newcommand{\deltaz}{\ensuremath{\Delta z}}
\newcommand{\deltay}{\ensuremath{\Delta y}}
\newcommand{\timeofflight}{\ensuremath{t_{\rm{tof}}}}	
\newcommand{\deltatof}{\ensuremath{\Delta \timeofflight}}

\newcommand{\RAA}{\ensuremath{R_{\rm{AA}}}}

\newcommand{\ncoll}{\ensuremath{N_{\rm{coll}}}}
\newcommand{\npart}{\ensuremath{N_{\rm{part}}}}

\newcommand{\dtacsum}{\ensuremath{\Delta t_{\mathrm{trig}}}}

\begin{document}

\begin{frontmatter}

\title{Measurement of inclusive \jpsi\ suppression in \AuAu\ collisions at \sqrtsNN\ = 200 GeV through the dimuon channel at STAR}

\author{
J.~Adam$^{13}$,
L.~Adamczyk$^{2}$,
J.~R.~Adams$^{37}$,
J.~K.~Adkins$^{28}$,
G.~Agakishiev$^{26}$,
M.~M.~Aggarwal$^{39}$,
Z.~Ahammed$^{59}$,
I.~Alekseev$^{3,33}$,
D.~M.~Anderson$^{53}$,
R.~Aoyama$^{56}$
% ...
}

% ...
\address{$^{2}$AGH University of Science and Technology, FPACS, Cracow 30-059, Poland}
\address{$^{3}$Alikhanov Institute for Theoretical and Experimental Physics, Moscow 117218, Russia}
% ...
\address{$^{13}$Creighton University, Omaha, Nebraska 68178}
% ...
\address{$^{26}$Joint Institute for Nuclear Research, Dubna 141 980, Russia}
% ...
\address{$^{28}$University of Kentucky, Lexington, Kentucky 40506-0055}
% ...
\address{$^{33}$National Research Nuclear University MEPhI, Moscow 115409, Russia}
% ...
\address{$^{37}$Ohio State University, Columbus, Ohio 43210}
% ...
\address{$^{39}$Panjab University, Chandigarh 160014, India}
% ...
\address{$^{53}$Texas A\&M University, College Station, Texas 77843}
% ...
\address{$^{56}$University of Tsukuba, Tsukuba, Ibaraki 305-8571, Japan}
% ...
\address{$^{59}$Variable Energy Cyclotron Centre, Kolkata 700064, India}
% ...

\begin{abstract}
\jpsi\ suppression has long been considered a sensitive signature of the formation of the Quark-Gluon Plasma (QGP) in relativistic heavy-ion collisions. In this letter, we present the first measurement of inclusive \jpsi\ production at mid-rapidity through the dimuon decay channel in \AuAu\ collisions at \mbox{\sqrtsNN\ = 200 GeV} with the STAR experiment. These measurements became possible after the installation of the Muon Telescope Detector was completed in 2014. The \jpsi\ yields are measured in a wide transverse momentum (\pT) range of 0.15 \gev\ to 12 \gev\ from central to peripheral collisions. They extend the kinematic reach of previous measurements at RHIC with improved precision. In the 0-10\% most central collisions, the \jpsi\ yield is suppressed by a factor of approximately 3 for $\pT>5$ \gev\ relative to that in \pp\ collisions scaled by the number of binary nucleon-nucleon collisions. The \jpsi\ nuclear modification factor displays little dependence on \pT\ in all centrality bins. Model calculations can qualitatively describe the data, providing further evidence for the color-screening effect experienced by \jpsi\ mesons in the QGP.
\end{abstract}


\end{frontmatter}


\section{Introduction}
Among the primary goals of high-energy heavy-ion physics are the creation of the Quark-Gluon Plasma (QGP) and the study of its properties \cite{Adams:2005dq}. These studies are being carried out at the Relativistic Heavy Ion Collider (RHIC) and the Large Hadron Collider (LHC). Among the various probes of the QGP, quarkonia play a special role as they are expected to dissociate in the medium when the Debye radius, inversely proportional to the medium temperature, becomes smaller than their size \cite{Matsui:1986dk}. Strong suppression of the \jpsi\ meson with respect to its yield in \pp\ collisions scaled by the number of binary nucleon-nucleon collisions has been observed at high transverse momenta (\pT) in central heavy-ion collisions at both RHIC and LHC energies \cite{Adamczyk:2013tvk, Adamczyk:2012ey, Adamczyk:2016srz, Adare:2006ns, Abelev:2013ila, Adam:2015rba, Khachatryan:2016ypw, Sirunyan:2017isk}. The level of suppression is beyond that expected from Cold Nuclear Matter (CNM) effects \cite{Adare:2007gn, Adam:2015jsa, Aaboud:2017cif}, which include modifications to the parton distribution function in nuclei \cite{Eskola:2009uj,Kovarik:2015cma}, nuclear absorption \cite{Gerschel:1988wn}, and radiative energy loss \cite{Arleo:2012hn}. This suggests that the reduction of the high-\pT\ \jpsi\ yield is, at least partially, due to the presence of the hot medium and the color-screening effect is believed to be the underlying mechanism. The real part of the $c\bar{c}$ potential can get color-screened statically in the medium \cite{Matsui:1986dk}, resulting in a broadening of the wave function, while the imaginary part of the potential is related to the dissociation of \jpsi\ arising from scattering with medium constituents. The latter is sometimes referred to as the dynamical color-screening effect or collisional dissociation \cite{Xu:1995eb,Yao:2018sgn,Sharma:2012dy}. Other effects have also been found to modify the observed \jpsi\ yield in heavy-ion collisions \cite{Andronic:2015wma}. A prominent contribution arises from the regeneration of \jpsi\ from deconfined charm and anti-charm quarks in the medium. It is responsible for the reduced suppression of low-\pT\ \jpsi's at the LHC compared to RHIC \cite{Abelev:2013ila} due to the larger charm production cross-section at the former. Also, the pre-resonance $c\bar{c}$ pairs in color-octet states could undergo energy loss in the medium before quarkonia are formed \cite{Sharma:2012dy}. Furthermore, significant feed-down contributions from excited charmonium states such as $\chi_{c}$ and $\psi$(2S) ($\sim$40\% \cite{Digal:2001ue}) as well as from b-hadron decays ($\sim$10-25\% above 5 \gev\ \cite{Adamczyk:2012ey}) add additional complications as the suppression level for mother particles in the medium could differ from that of directly produced \jpsi, i.e. ones not from decays. Model calculations, incorporating either continuous dissociation and regeneration throughout the medium evolution \cite{Yan:2006ve, Zhou:2014kka, Zhao:2010nk, Zhao:2011cv} or a complete melting of all \jpsi\ above the dissociation temperature and regeneration at the phase boundary \cite{BraunMunzinger:2000px, Andronic:2017pug} or collisional dissociation plus energy loss \cite{Sharma:2012dy}, can qualitatively describe the experimental measurements. To provide further constraints on models and ultimately help infer the medium temperature, detailed differential measurements of \jpsi\ suppression over a broad kinematic range with good precision are needed since the aforementioned effects depend on the momentum of the \jpsi\ as well as the collision geometry. Measurements through the dimuon decay channel are preferred compared to the dielectron channel because of the greatly reduced multiple scattering in the material and negligible bremsstrahlung.

In this letter, we present a new measurement of \jpsi\ suppression at mid-rapidity in \AuAu\ collisions at \mbox{\sqrtsNN\ = 200 GeV} through the dimuon decay channel by the Solenoidal Tracker At RHIC (STAR) experiment \cite{Ackermann:2002ad}. The inclusive \jpsi\ sample used in this analysis includes decays from excited charmonia and \mbox{b-hadrons}. This measurement is made possible by the Muon Telescope Detector (MTD) designed for triggering on and identifying muons \cite{Ruan:2009ug}, which was completed in early 2014. Compared to previous mid-rapidity measurements through the dielectron channel at RHIC \cite{Adamczyk:2013tvk, Adamczyk:2012ey, Adamczyk:2016srz, Adare:2006ns}, the new results extend the kinematic reach towards high \pT\ with better precision. 

\section{Experiment, dataset and analysis}
The data sample used in this analysis was collected from \AuAu\ collisions at \mbox{\sqrtsNN\ = 200 GeV} in 2014. Events were selected by a dedicated dimuon trigger, which requires at least two muon signals accepted by the MTD in coincidence with signals in the Zero Degree Calorimeters (ZDCs) \cite{Adler:2000bd}. The MTD consists of 122 modules made from multi-gap resistive plate chambers, providing timing information for particles passing through. It resides outside of the solenoid magnet at a radius of 403 cm, and covers about 45\% in azimuth ($\varphi$) within the pseudo-rapidity range of $|\eta| < 0.5$. The magnet also acts as a hadron absorber amounting to 5 interaction lengths. Using double-ended readout strips, the timing resolution of the MTD is about 100 ps, and the intrinsic spatial resolutions are 1.4 cm and 0.9 cm in $r\varphi$ and beam ($z$) directions, respectively \cite{Yang:2014xta}. Variable numbers of MTD modules, ranging from 2 to 5 and located at the same $\eta$, are grouped into 28 trigger patches. The earliest signal in each trigger patch is picked up and accepted by the trigger system if its flight time (\dtacsum) falls into a pre-defined online trigger time window. The \dtacsum\ is the difference in time measured by the MTD and the start time provided by the Vertex Position Detector (VPD), which is a fast detector covering $4.24<|\eta|<5.1$ \cite{Llope:2014nva}. In total, an integrated luminosity of 14.2 \invnb\ was sampled by the dimuon trigger. 

The main tracking device is the Time Projection Chamber (TPC) \cite{Anderson:2003ur} immersed in a solenoidal magnetic field of 0.5 T and covering full azimuth within $|\eta|<1.0$. The primary event vertex is reconstructed using TPC tracks, and required to be within $\pm$100 cm to the center of STAR along the beam line and within 1.8 cm in radial direction. To reject pileup events, the vertex positions determined by the TPC and the VPD are required to agree within 3 cm along the beam direction. The collision centrality is determined by matching the multiplicity distribution of charged tracks from data to the Monte Carlo Glauber model \cite{Abelev:2008ab}. The selected charged tracks are within $|\eta|<0.5$ and have Distances of Closest Approach (DCA) to the primary vertex of less than 3 cm.

\section{Muon identification and \jpsi\ signal}
Since particles of low momenta are mostly absorbed in the material in front of the MTD, only charged tracks with $\pT>1.3$ \gev\ are accepted. To assure high quality, the number of TPC space points used for track reconstruction is required to be no less than 15. The ratio of the number of used to the maximum possible number of TPC space points is required to be larger than 0.52 in order to reject split tracks. Furthermore, a track's DCA to the primary vertex needs to be smaller than 1 cm to get accepted. It is then refit including the primary vertex to improve the momentum resolution. To identify muon candidates, the specific energy loss ($dE/dx$) measured in the TPC, quantified as \nsigmapi, is used:
\begin{equation}
\nsigmapi=\frac{\ln(dE/dx)_{\rm{measured}}-\ln(dE/dx)_{\rm{theory}}^{\pi}}{\sigma(\ln(dE/dx))}
\end{equation}
Here $(dE/dx)_{\rm{measured}}$ is the measured energy loss in the TPC, $(dE/dx)_{\rm{theory}}^{\pi}$ is the expected energy loss for a pion based on the Bichsel formalism \cite{Bichsel:2006cs} and $\sigma(\ln(dE/dx))$ stands for the resolution of the $\ln(dE/dx)$ measurement. Since muons lose more energy per unit of path length by about half of the $dE/dx$ resolution than pions, an asymmetric cut of $-1<\nsigmapi<3$ is used. 

To take advantage of the MTD, tracks are propagated from the outermost TPC space points to the MTD and matched to the closest MTD hits found within a large search window. The propagation is done assuming the track is a muon. If more than one track is matched to the same hit, the closest track is chosen. The track propagation is based on a helix model taking into account both the variation of the magnetic field and the energy loss along the trajectory. The magnetic field changes from -0.5 T in the TPC  to +1.26 T in the magnet steel, and vanishes outside of the magnet. The average energy loss in the material is parametrized using the GEANT3 \cite{Brun:1987ma} simulation of the STAR detector. Once a track-hit association is established, requirements on the distance between the MTD hit position and projected track position are applied to further reject hadrons. In the local coordinate frame of the MTD module where the associated hit resides, differences in both {\it y} and {\it z} directions, i.e. \deltay\ and \deltaz, are required to be less than 2(2.5)$\sigma_{\deltay,\deltaz}$ for tracks with \pT\ $<$($\ge$) 3 \gev, where $\sigma_{\deltay,\deltaz}$ stand for the \pT-dependent \deltay\ and \deltaz\ resolutions. The {\it y} and {\it z} directions in the local coordinate frame correspond to the azimuthal and {\it z} directions in the global coordinate frame, respectively. Since muons reach the MTD faster than background hadrons, the time of flight (\timeofflight) of a particle measured by the MTD with respect to the start time provided by the VPD should be within 0.75 ns of the expected flight time extracted from the track propagation, i.e. $\deltatof<0.75$ ns. A summary of the muon PID cuts is listed in Table \ref{tab:MuonPID}.
\begin{table}[htbp]
\centering
\begin{tabular}{c|c} 
 \hline
Detector used & Muon PID cuts \\
 \hline
TPC & $-1<\nsigmapi<3$\\
 \hline
 \multirow{3}{*}{MTD} & $|\deltay|<2(2.5)\sigma_{\deltay}$ for \pT\ $<$($\ge$) 3 \gev\ \\
 & $|\deltaz|<2(2.5)\sigma_{\deltaz}$ for \pT\ $<$($\ge$) 3 \gev\ \\
 & $\deltatof<0.75$ ns \\
 \hline
\end{tabular}
\caption{List of muon PID cuts. }
\label{tab:MuonPID}
\end{table}

\begin{figure}[htbp]
\begin{minipage}{0.49\linewidth}
\centerline{\includegraphics[width=0.95\linewidth]{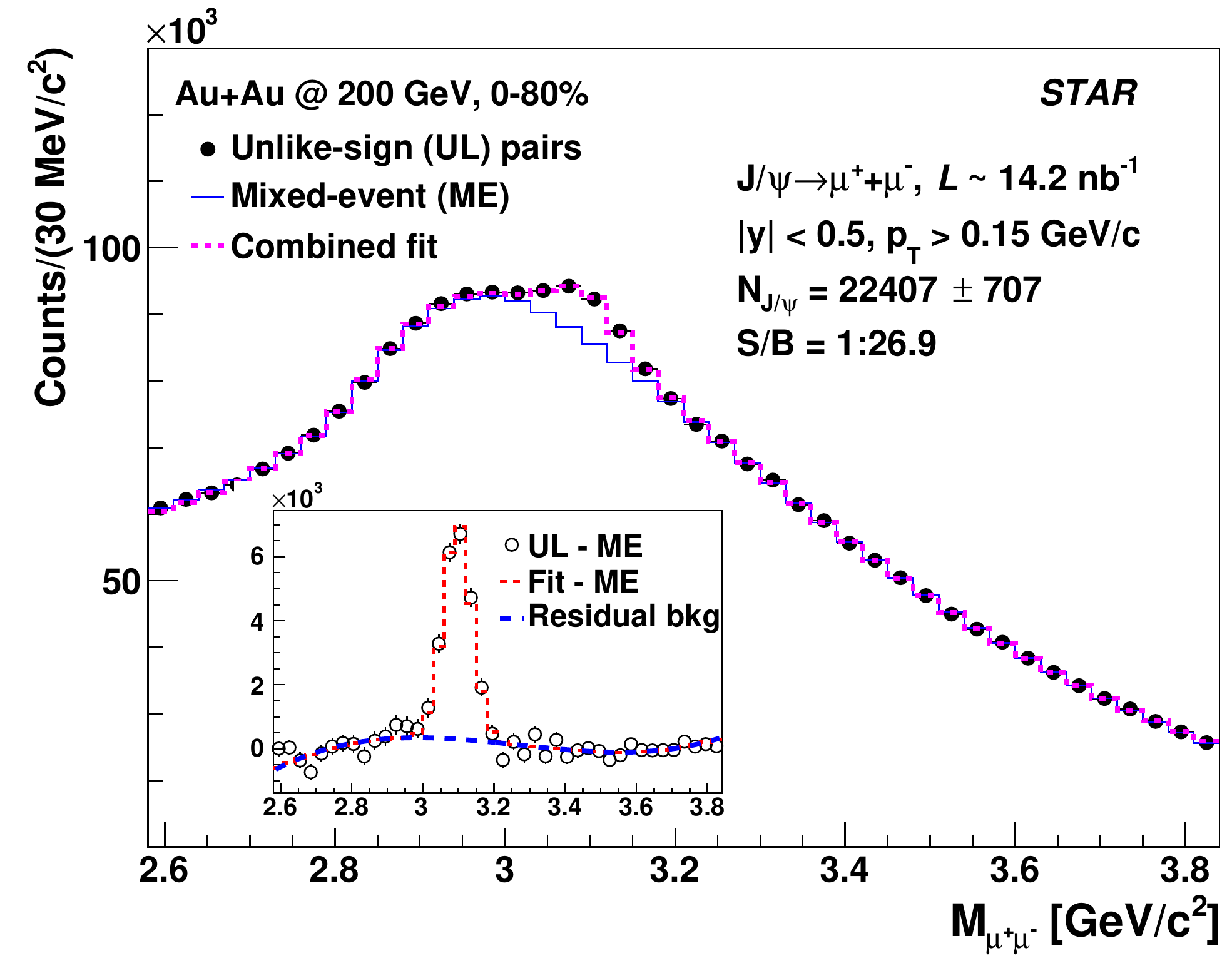}}
\end{minipage}
\begin{minipage}{0.49\linewidth}
\centerline{\includegraphics[width=0.95\linewidth]{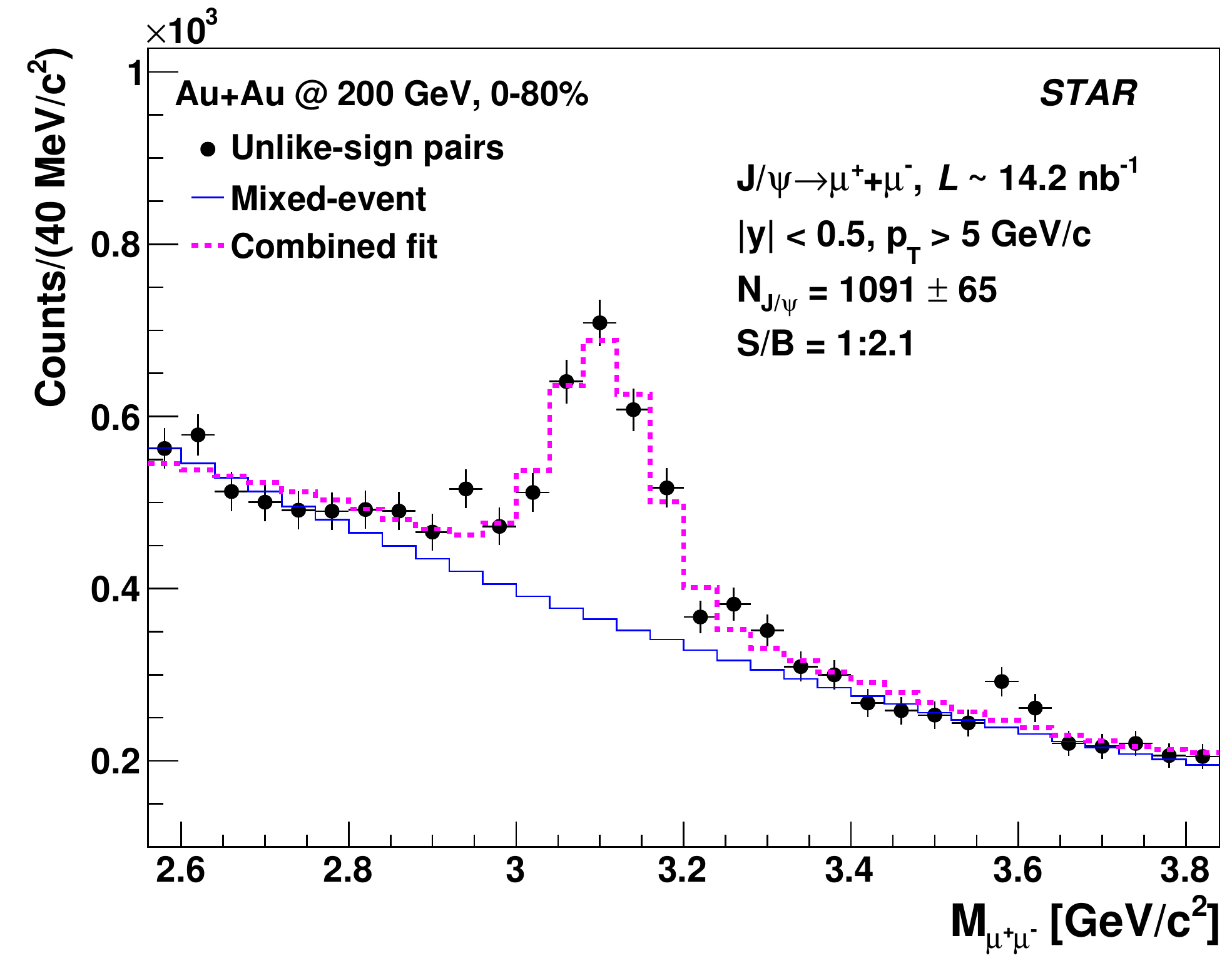}}
\end{minipage}
\caption[]{Invariant mass distributions of unlike-sign pairs with \pT\ above 0.15 (left) and 5 (right) \gev\ in 0-80\% central \AuAu\ collisions. The same-event (filled circles) and mixed-event (blue histogram) distributions are shown along with the combined fit (signal+background). The insert in the left panel shows the signal distribution (open circles) and combined fit (dashed line), with the combinatorial background subtracted, superimposed with the residual background (long dashed line). }
\label{fig:InvMass}
\end{figure}

Unlike-sign muon candidates from the same event are paired to reconstruct the invariant mass of \jpsi\ signals, examples of which are shown in Fig. \ref{fig:InvMass} (filled circles) for pair \pT\ above 0.15 and 5 \gev, respectively, in 0-80\% \AuAu\ collisions. \jpsi\ candidates with $\pT <0.15$ \gev\ are excluded to avoid the influence of the very low-\pT\ \jpsi's likely originating from coherent photoproduction \cite{Adam:2015gba}.
Also shown in the figure are the scaled unlike-sign distributions from the mixed events (histogram), which have good statistical precision and same acceptance as the signal pairs, as estimates of the combinatorial background. The mixed-event distributions are constructed by pairing muon candidates of opposite charge signs in different events. To ensure proper and sufficient mixing, each event is mixed with 100 other events in the same category, i.e. 16 bins in 0-80\% centrality, 20 bins in vertex $|z|< $100 cm, and 24 bins in the reconstructed event plane angle \cite{Poskanzer:1998yz}. The normalization factors for the mixed-event unlike-sign distributions are determined by a linear fit to the ratio of like-sign distributions from same and mixed events within the mass window of 2.7 to 3.8 \gevtwo. The same-event unlike-sign distributions are fitted using the maximum likelihood method with three components: i) a Gaussian function representing the \jpsi\ signal, ii) the mixed-event distribution for the combinatorial background, and iii) a first-order or third-order polynomial function describing the residual background. The raw \jpsi\ counts are obtained directly from the fits. The widths of the Gaussian distributions in different \jpsi\ \pT\ bins are fixed according to the detector simulations tuned to match the data, while the third-order polynomial function is only used at low \pT\ in central collisions. The combined fits are also shown in Fig. \ref{fig:InvMass} as dashed lines, from which the extracted number of \jpsi\ above 0.15 \gev\ is 22407 with a significance of 31.7 within the 0-80\% centrality bin.

\section{Efficiency correction}
Corrections for signal reconstruction efficiency and detector acceptance are evaluated using a combination of detector simulation and data-driven methods. They include efficiencies for the TPC tracking, MTD matching, particle identification (PID) of muons, and MTD triggering. 

The TPC tracking efficiency, including the TPC acceptance, is evaluated by embedding simulated \jpsi\ signals into real events. The input \jpsi's, weighted with previously published \pT\ distributions \cite{Adamczyk:2013tvk, Adamczyk:2012ey}, are forced to decay into two muons and then passed through the GEANT3 detector simulations. The simulation signals are digitized and embedded into real data, and the same reconstruction procedure as for the real data is applied. Since the TPC tracking efficiency depends strongly on the occupancy, the number of embedded \jpsi\ is set to be 5\% of the event multiplicity to avoid any significant distortion to the TPC performance. Additional correction factors are applied to account for the different vertex distributions between data and embedding samples, as well as additional luminosity and centrality dependences of the TPC inefficiency in local areas which are not accounted for in the embedding.

For the matching efficiency between TPC tracks and MTD hits, the MTD acceptance is modeled in the detector simulation, whereas the in-situ response of each MTD module is obtained from cosmic ray data. For modules located in the bottom hemisphere of the STAR detector, the response efficiency in each module is parametrized. For those in the top hemisphere, tracks travel in opposite direction to those in collision data. This, combined with the energy loss effect in the material, results in incorrect efficiency estimates at low \pT. Therefore, the \pT\ dependence of the average response efficiency for all the bottom modules is used as a template for top modules, while the absolute scale is determined according to the response efficiency of each individual top module above 5 \gev, where the efficiency reaches a plateau. 

\begin{figure}[htbp]
\begin{minipage}{0.95\linewidth}
\centerline{\includegraphics[width=0.6\linewidth]{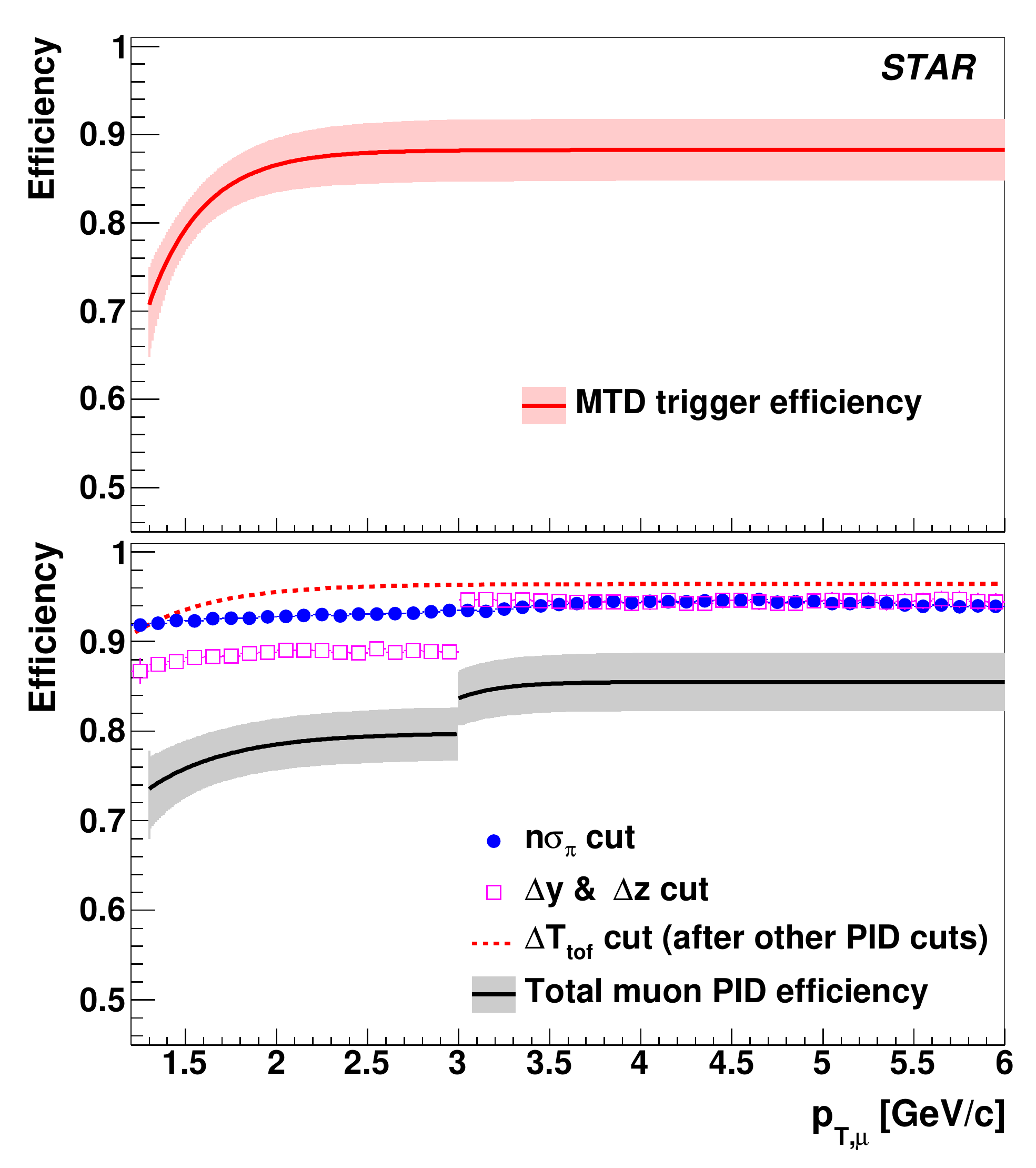}}
\end{minipage}
\caption[]{Efficiencies for MTD triggering (upper) and various muon PID cuts (lower) as a function of muon \pT. The bands represent the systematic uncertainties.}
\label{fig:MtdEff}
\end{figure}

The efficiencies related to muon identification cuts on \nsigmapi, \deltay, \deltaz\ are extracted from embedding, while the \deltatof\ efficiency is evaluated with the ``tag-and-probe" method using real data since the timing information is not simulated. In this approach, a ``tag" muon and a ``probe" muon are paired to construct the \jpsi\ signal. The tag muons are always selected with strict PID cuts in order to increase the signal-to-background ratio, while the probe muons are selected with the standard \nsigmapi, \deltay, \deltaz\ cuts as well as two cases of the \deltatof\ cut, i.e. no \deltatof\ cut and $\deltatof<0.75$ ns. The ratio of the \jpsi\ yields from the two cases as a function of the probe muon \pT\ is parametrized as the \deltatof\ cut efficiency for muons. Figure \ref{fig:MtdEff}, lower panel, shows the \nsigmapi, \deltay\ plus \deltaz, and \deltatof\ cut efficiencies as well as the combined muon PID efficiency as a function of muon \pT. The discontinuity at 3 \gev\ is due to the change in the \deltay\ and \deltaz\ cuts (see Table \ref{tab:MuonPID}). The muon PID efficiency is about 73\% at $\pT=1.3$ \gev, and reaches a plateau of about 85\% at high \pT.

The MTD trigger efficiency originates from the requirement that an MTD signal is accepted for triggering only when its flight time, \dtacsum, falls within a pre-defined trigger time window. Since the MTD timing information is not simulated, the \pp\ data taken in 2015 are used instead to estimate the MTD trigger efficiency. This approach is possible because: i) the trigger efficiency in \pp\ data is very close to 100\% due to the very loose trigger time window used during data-taking, ii) the MTD trigger system stayed the same between 2014 and 2015, iii) occupancy of the MTD system is very small, e.g. there are on average 6.2 hits in 122 modules in 0-10\% central \AuAu\ collisions, and therefore the multiplicity difference between \pp\ and \AuAu\ collisions does not play a role. In \pp\ collisions, the \dtacsum\ distributions for pure muons are obtained by statistically subtracting the \dtacsum\ distributions for like-sign muon pairs from those for unlike-sign pairs in the mass window of \mbox{[3.0, 3.2] \gevtwo}. The resulting \dtacsum\ distributions are then shifted to account for the difference in the mean values of \dtacsum\ distributions between \pp\ and \AuAu\ data, arising from the different global timing setup during the online data taking. The different VPD resolutions in \pp\ and \AuAu\ collisions due to different multiplicities are determined from online trigger data and taken into account. Furthermore, the relative abundances of the trigger signals in each trigger patch are obtained from \AuAu\ data and used to weight the \dtacsum\ distributions of the corresponding trigger patches in the \pp\ data. The resulting trigger efficiency for 2014 \AuAu\ data is shown in the upper panel of Fig. \ref{fig:MtdEff}. It increases from 71\% at lowest \pT\ to 88\% at high \pT.

In order to extract the total \jpsi\ reconstruction efficiency, single muon efficiencies determined from data are applied to the \jpsi\ simulation, which takes the decay kinematics properly into account using the PYTHIA event generator \cite{Sjostrand:2006za}. 

\section{Systematic uncertainties}
\paragraph{Signal extraction} Variations are made to different aspects of the signal extraction procedure and the maximum differences from the default values are taken as the systematic uncertainties. When obtaining the normalization factors for the mixed-event background, the fit range is varied and the fit function is changed from first-order to zeroth-order polynomial. To extract the raw \jpsi\ counts, the binning of the invariant mass distributions is changed, and so is the fit range. Different functional forms, such as a Crystal-ball function \cite{Gaiser:1982yw} and line-shapes from tuned simulation, are used for signal shape, while polynomial functions of different orders are substituted for describing the residual background. Finally, the bin-counting method, with the residual background contribution removed, is tried. 

\paragraph{TPC tracking} The uncertainties in the TPC tracking efficiency are evaluated by changing the track quality cuts simultaneously in the data analysis and in extracting the tracking efficiency from the embedding sample, and repeating the whole procedure to obtain the corrected \jpsi\ yields. The maximum differences from the default case are seen to be almost independent of \jpsi\ \pT\ for 0-80\% centrality bin. A constant fit gives a \pT-independent uncertainty of 5.8\%. For finer centrality bins the same uncertainty is used, which covers most of the variation seen in these centrality bins. Furthermore, an overall 2\% uncertainty is assigned for the correction factor used to account for the mismatch of the vertex distributions between data and embedding. An uncertainty of 0.2\%-6.1\% from central to peripheral events is associated with the correction of the luminosity and centrality dependent TPC inefficiencies. An additional 5\% overall uncertainty is assigned based on the comparison of the like-sign muon pair yields in different luminosity profiles. 

\paragraph{MTD matching efficiency} Two sources of uncertainties in the MTD response efficiency are investigated. Firstly, the statistical errors on the cosmic ray data, used in determining the efficiency curves for the bottom modules and the scale factors for top modules, are treated as a source of uncertainty. The 68\% confidence intervals of the fit results are taken. Their influence on the \jpsi\ spectrum is assessed by randomizing the efficiency curves of each module independently within their respective errors many times and checking the spread of the resulting response efficiency as a function of \jpsi\ \pT. Secondly, the uncertainty from the assumption of using the efficiency template for top modules is estimated by taking the average absolute difference between the response efficiency curves of bottom modules and the template efficiency. Furthermore, the MTD matching efficiencies extracted from simulation and from cosmic ray data are compared and the difference is taken as an additional source of uncertainty. The total uncertainty on the MTD matching efficiency is taken as the quadratic sum of these three sources. It is 9.1\% for $0.15<\pT<1$ \gev\ and decreases to 1.0\% at the highest \pT\ bin. 

\paragraph{Muon PID} The uncertainties in the \nsigmapi, \deltay, \deltaz\ cut efficiencies, extracted from the embedding sample, are estimated the same way as the TPC tracking efficiency. For the \deltatof\ cut efficiency, the uncertainty comes mainly from the statistical errors on the data points used to extract this efficiency. It is evaluated by randomly changing the data points independently within their individual errors, fitting the randomized data points, and taking the root-mean-square of the resulting efficiency distributions in each muon \pT\ bin. The total uncertainty on the muon PID efficiency, shown as the band around the efficiency curve in the lower panel of Fig. \ref{fig:MtdEff}, is the quadratic sum of the two contributions. 

\paragraph{MTD triggering} The uncertainty in the MTD trigger efficiency is shown as the shaded band around the efficiency curve in the upper panel of Fig. \ref{fig:MtdEff}. The main contributions to the uncertainties arise from the procedure of using \dtacsum\ distributions from \pp\ data to extract this efficiency for \AuAu\ analysis. The residual difference in the mean values of the \dtacsum\ distributions for \jpsi-decayed muons from the \pp\ data after shifting and the \AuAu\ data is taken into account, and so are the uncertainties in the extracted widths of the \dtacsum\ distributions as a function of muon \pT. Alternatively, the widths of the \dtacsum\ distributions in \AuAu\ collisions are calculated based on those from \pp\ collisions combined with the difference in the VPD resolutions between the two collision systems. The maximum deviations of all these variations in each muon \pT\ bin are fit to obtain the uncertainty curves. The statistical errors on the \pp\ data are also included. Furthermore, the MTD trigger efficiency is verified by comparing the results for muon candidates from the 2015 \pp\ data and the 2014 minimum-bias \AuAu\ data, and the difference is assigned as an additional systematic uncertainty which varies from 3.9\% at low \pT\ to 1.1\% at high \pT.

All the aforementioned uncertainties are listed in Table \ref{tab:SysUncert} for two representative \jpsi\ \pT\ bins, i.e. $0.15<\pT<1$ \gev\ and $5<\pT<6$ \gev\ in the 0-80\% centrality class.
\begin{table}[htbp]
\centering
\begin{tabular}{c|c|c|c} 
 \hline
 \hline
Uncertainty sources & $0.15<\pT<1$ \gev\ & $5<\pT<6$ \gev\ & Correlation \\
 \hline
Signal extraction & 5.2\% & 2.2\% & No\\
\hline
TPC tracking & 7.9\% & 7.9\% & Fully\\
\hline
\multirow{2}{*}{MTD matching} & 1.5\% &  0.3\% & No \\
& 9.0\% & 5.8\% & Fully\\
\hline
\multirow{2}{*}{Muon PID}& 5.0\% & 3.0\% & No\\
& 8.5\% & 5.9\% & Fully\\
\hline
\multirow{2}{*}{MTD triggering} & 1.6\% & 1.7\% & No\\
& 7.0\% & 7.5\% & Fully \\
\hline\hline
Total & 18.0\% & 14.3\% & Largely \\
  \hline
\end{tabular}
\caption{Individual and total systematic uncertainties for two representative \jpsi\ \pT\ bins, i.e. $0.15<\pT<1$ \gev\ and $5<\pT<6$ \gev\ in 0-80\% centrality class. Correlated and uncorrelated components of various uncertainties are separated.}
\label{tab:SysUncert}
\end{table}
The total uncertainties are the quadratic sum of all the individual sources. The uncertainties are fully or largely correlated across different \pT\ and centrality bins except for the signal extraction uncertainty.

\section{Results and Discussion}
\begin{figure}[htbp]
\begin{minipage}{0.95\linewidth}
\centerline{\includegraphics[width=0.6\linewidth]{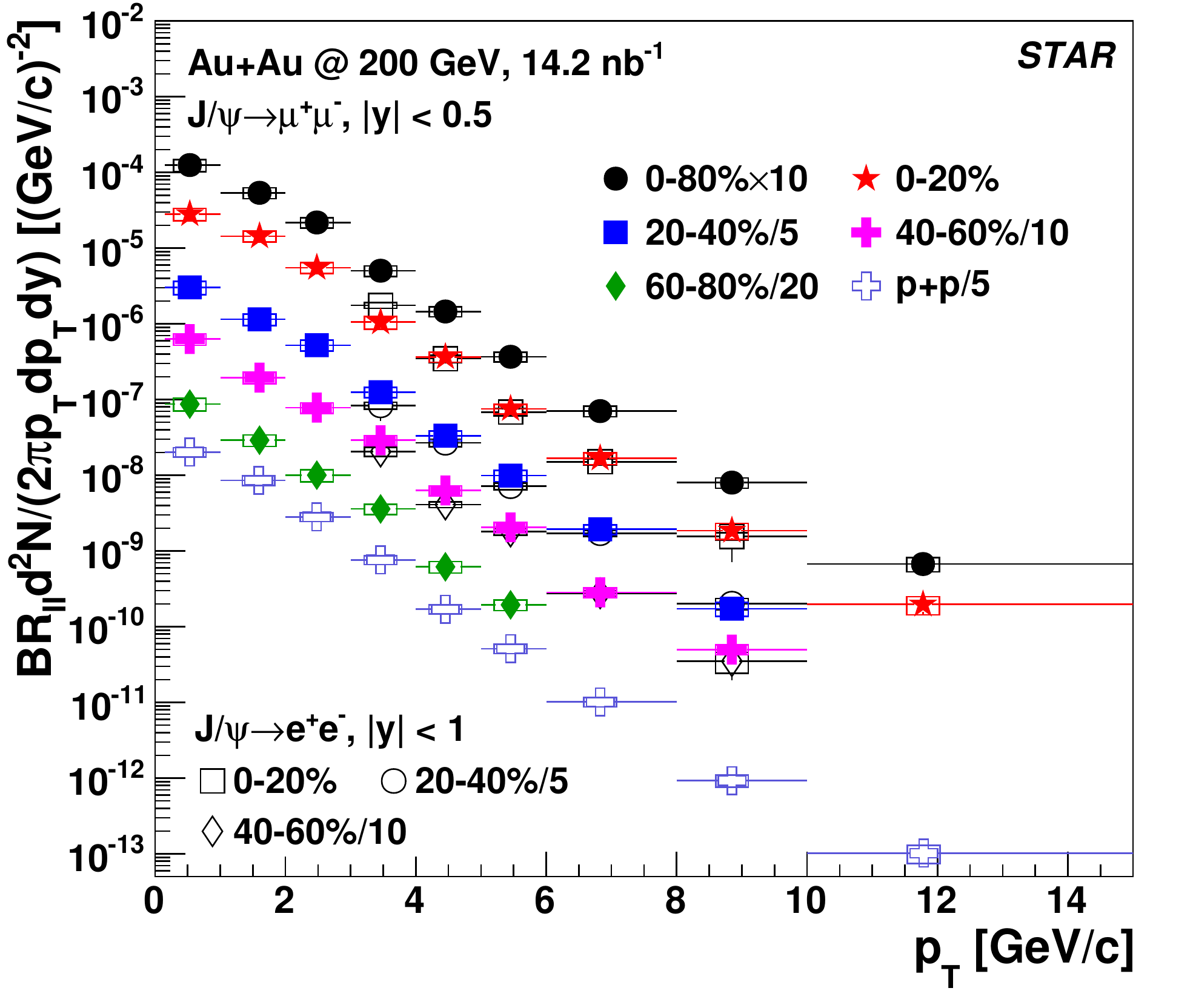}}
\end{minipage}
\caption[]{Invariant yields of inclusive \jpsi, measured through their respective decay channels, in different centrality intervals of \AuAu\ collisions at \mbox{\sqrtsNN\ = 200 GeV}. The reference distribution from \pp\ collisions is also shown. The vertical error bars and boxes around the data points represent the statistical errors and systematic uncertainties, respectively. The horizontal error bars indicate the bin widths. In most cases, the statistical error bars are smaller than the marker size. Multiplicative factors are applied to the spectra for clarity.}
\label{fig:JpsiinvYield}
\end{figure}

The invariant yields of inclusive \jpsi\ within $|y| < 0.5$ as a function of \pT, measured through the dimuon channel, are shown in Fig. \ref{fig:JpsiinvYield} as filled symbols for five different centrality intervals in \AuAu\ collisions at \mbox{\sqrtsNN\ = 200 GeV}. 
The data points are placed at the \pT\ positions whose yields are equal to the average yields of the bins \cite{Lafferty:1994cj}. They are determined from fitting the yields iteratively with an empirical functional form of \mbox{$A\times\pT\times(1+(\pT/B)^{2})^{-C}$}, where $A, B$, and $C$ are free parameters. The bin widths are indicated by the horizontal error bars around the data points. Also shown in Fig. \ref{fig:JpsiinvYield}, depicted by open symbols, are the updated invariant yields of inclusive \jpsi\ within $|y| < 1.0$ measured through the dielectron channel. In the original paper [6], incorrect values were used for the \jpsi\ reconstruction efficiency. A re-analysis showed that the originally used efficiencies were too small by 21\% (9\%) in 0-20\% (40-60\%) centrality class with little \pT\ dependence.

The modification of \jpsi\ production is quantified using the nuclear modification factor (\RAA):
\begin{equation}
\RAA=\frac{1}{\langle\ncoll\rangle}\times\frac{(\frac{d^{2}N_{\jpsi}}{2\pi\pT d\pT dy})_{\rm{\AuAu}}}{(\frac{d^{2}N_{\jpsi}}{2\pi\pT d\pT dy})_{p+p}}
\end{equation}
where $\langle\ncoll\rangle$ is the average number of binary nucleon-nucleon collisions in a given centrality bin and $(\frac{d^{2}N_{\jpsi}}{2\pi\pT d\pT dy})_{\rm{\AuAu}}$, $(\frac{d^{2}N_{\jpsi}}{2\pi\pT d\pT dy})_{p+p}$ are the invariant \jpsi\ yields in \AuAu\ and \pp\ collisions, respectively. The uncertainty on \ncoll, evaluated by changing various parameters in the Glauber model, increases from 2.8\% in the 0-10\% most central collisions to 45\% in the 70-80\% most peripheral collisions. The reference \jpsi\ distribution in \pp\ collisions at \mbox{\sqrts\ = 200 GeV} is obtained by combining STAR and PHENIX measurements \cite{Adam:2018jmp,Adare:2009js}, and shown as open crosses in Fig. \ref{fig:JpsiinvYield}. The systematic uncertainties, shown as boxes around data points, include the 10\% global uncertainty and are largely correlated between \pT\ bins.

\begin{figure}[htbp]
\begin{minipage}{0.98\linewidth}
\centerline{\includegraphics[width=1.0\linewidth]{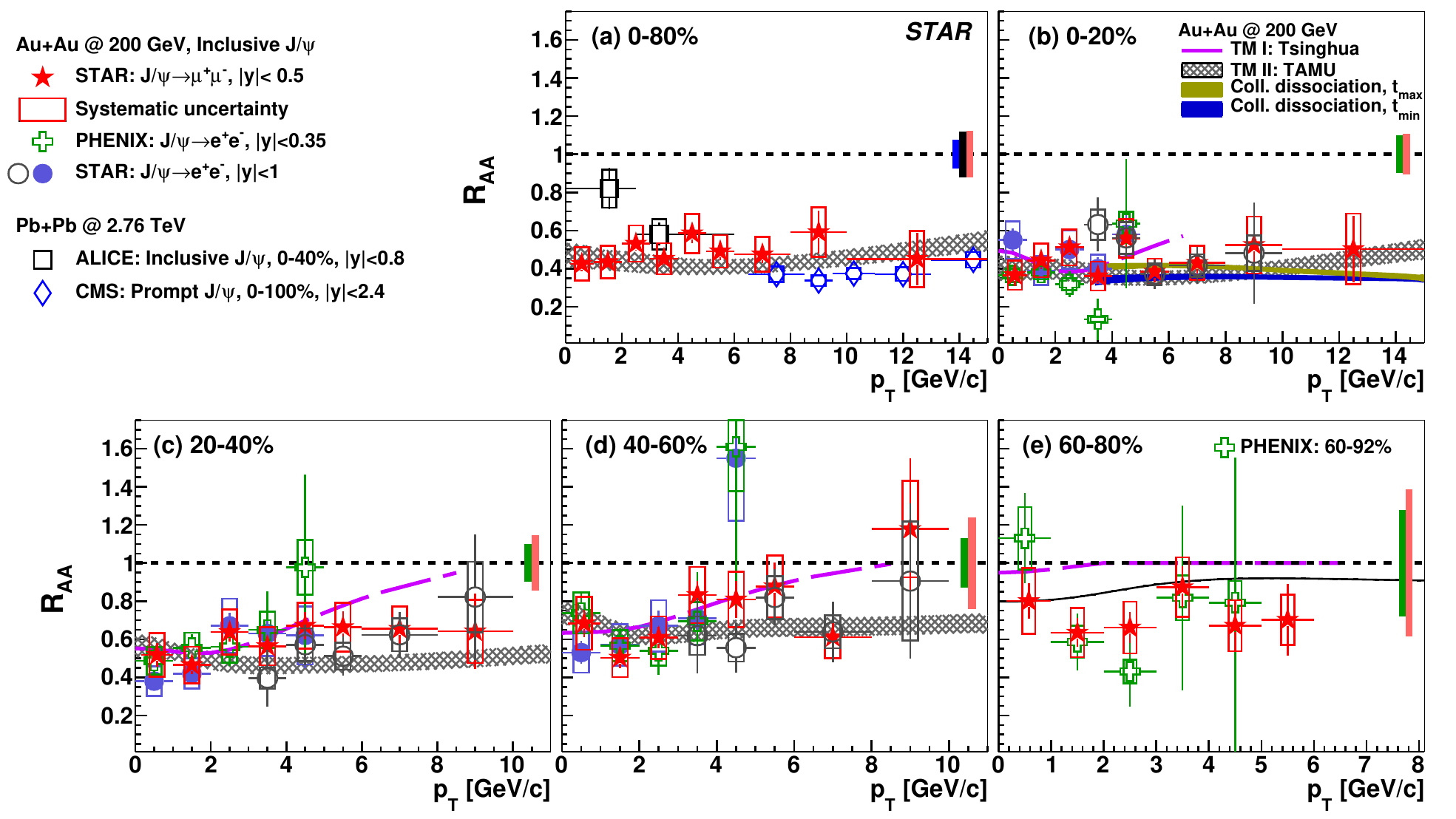}}
\end{minipage}
\caption[]{\jpsi\ \RAA\ as a function of \pT\ in different centrality intervals of 200 GeV \AuAu\ collisions. The error bars and boxes around data points represent the total statistical errors and systematic uncertainties from both \AuAu\ and \pp\ measurements. The boxes at unity show the global uncertainties, which for this analysis include the 10\% global uncertainty on \pp\ reference and the \ncoll\ uncertainties. Other measurements \cite{Adamczyk:2013tvk, Adamczyk:2012ey, Adare:2006ns,Adam:2015rba,Khachatryan:2016ypw} and model calculations \cite{Yan:2006ve, Zhou:2014kka, Zhao:2010nk,Sharma:2012dy} are shown for comparison. In the 0-80\% panel, the boxes at unity from left to right correspond to CMS, ALICE and STAR results, while for other panels the left band is for PHENIX and the right one for STAR.}
\label{fig:JpsiRaaVsPt}
\end{figure}

Figure \ref{fig:JpsiRaaVsPt} shows \jpsi\ \RAA\ as a function of \pT\ in 200 GeV Au+Au collisions as filled stars.
In all centrality bins, the \jpsi\ production is suppressed at low \pT, which is likely due to the combination of the CNM effects \cite{Adare:2007gn} and the dissociation in the QGP. In the 60-80\% centrality bin, the normalization uncertainty is large. Within the current uncertainties, the \jpsi\ \RAA\ shows little dependence on \pT. A sizable suppression in the \jpsi\ yield is present up to the largest measured \pT\ bin in central and semi-central collisions. There are several effects that could influence the \pT\ dependence of \RAA. The CNM effects decrease with increasing \pT. High \pT\ \jpsi's spend less time in the medium and are therefore less likely to be dissociated \cite{Blaizot:1988ec, Karsch:1987zw}. Furthermore, the relative contributions from b-hadron decays, whose suppression level is expected to be smaller than that of direct \jpsi\ \cite{Zhao:2010nk}, rise with increasing \pT. Also shown in Fig. 4 as open circles are the updated \jpsi\ \RAA\ measured through the dielectron channel. These values have been recalculated using the updated \jpsi\ yields in \AuAu\ collisions shown in Fig. \ref{fig:JpsiinvYield} and the new \pp\ reference derived in this analysis. Compared to the previously published results on \jpsi\ \RAA\ \cite{Adamczyk:2013tvk, Adamczyk:2012ey, Adare:2006ns}, the new results have better precision and span a wider kinematic range. In the overlapping range, good agreement is seen. In the upper left panel of Fig. \ref{fig:JpsiRaaVsPt}, similar measurements in \PbPb\ collisions at \mbox{\sqrtsNN\ = 2.76 TeV} are shown below \mbox{6 \gev} for inclusive \jpsi\ in 0-40\% centrality \cite{Adam:2015rba} and between \mbox{6.5-15 \gev} for prompt \jpsi\ in 0-100\% centrality \cite{Khachatryan:2016ypw}. The \jpsi\ \RAA\ measured in 0-80\% \AuAu\ collisions at  \mbox{\sqrtsNN\ = 200 GeV} is substantially below that at the LHC at low \pT, but systematically larger at higher \pT. Shown as long dashed lines and shaded areas are two transport model calculations for 200 GeV \AuAu\ collisions from Tsinghua \cite{Liu:2009nb} and TAMU \cite{Zhao:2010nk} groups, which take into account dissociation and regeneration contributions. The Tsinghua model agrees reasonably well with data at low \pT, but shows a different trend at high \pT. On the other hand, the TAMU model gives a fairly good description of data even though the central values of data points are mainly at the upper limit of the model calculation from intermediate to high \pT\ in non-peripheral events. In 0-20\% centrality, the two solid bands extending from 3.5 to 15 GeV/c are theoretical calculations for \AuAu\ collisions with two different values of the \jpsi\ formation time \cite{Sharma:2012dy}. This calculation uses vacuum \jpsi\ wave function without any screening effect, and includes both radiative energy loss of color-octet $c\bar{c}$ pairs and collisional dissociation of \jpsi. The regeneration is ignored as its contribution is small at large \pT. Both scenarios are consistent with data. All these model calculations include feed-down contributions from excited charmonia and b-hadron decays, as well as CNM effects.

\begin{figure}[htbp]
\begin{minipage}{0.49\linewidth}
\centerline{\includegraphics[width=1.0\linewidth]{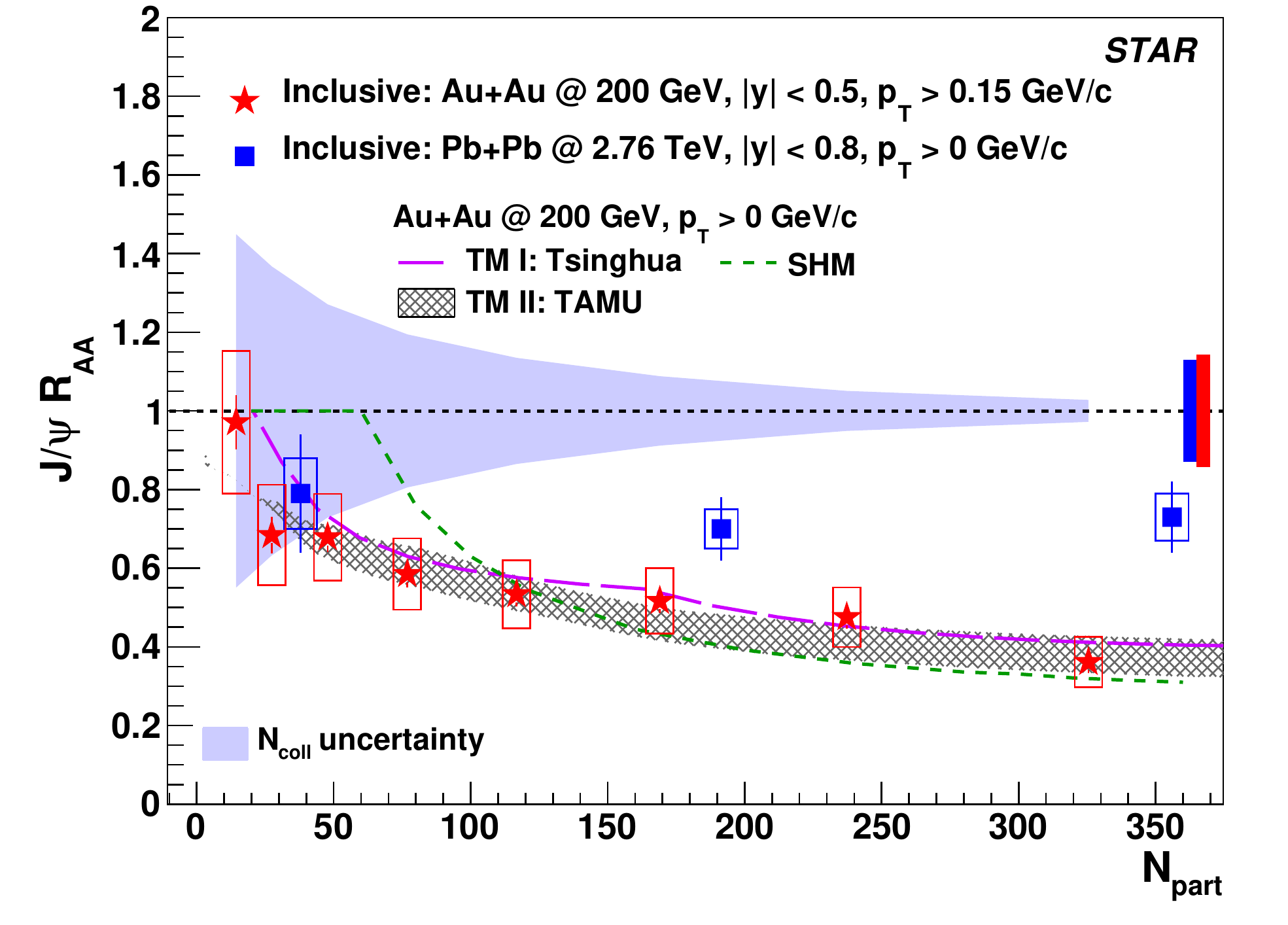}}
\end{minipage}
\begin{minipage}{0.49\linewidth}
\centerline{\includegraphics[width=1.0\linewidth]{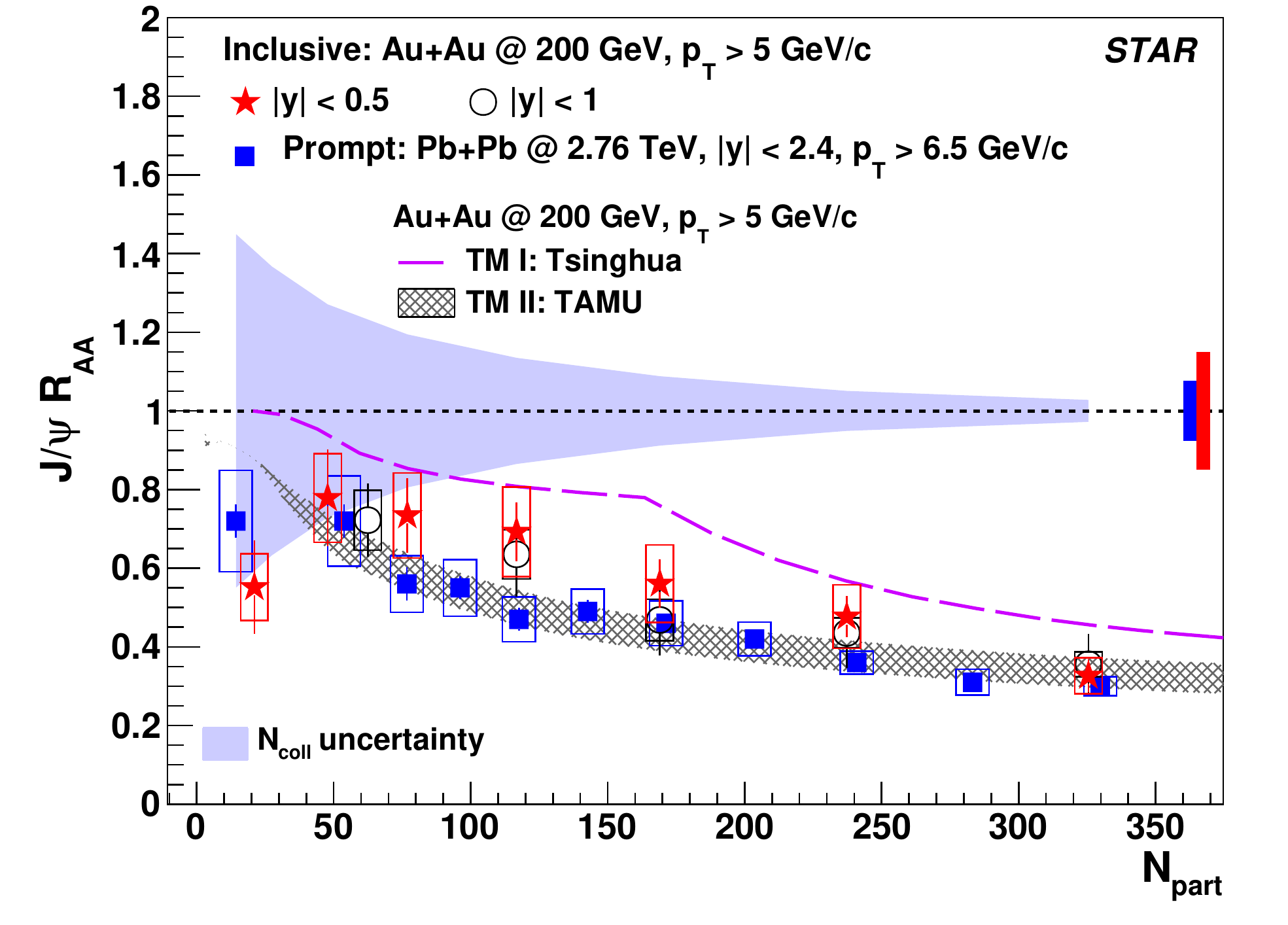}}
\end{minipage}
\caption[]{\jpsi\ \RAA\ as a function of \npart\ above 0.15 and 5 \gev\ in \AuAu\ collisions at \mbox{\sqrtsNN\ = 200 GeV}, compared to those for \PbPb\ collisions at \mbox{\sqrtsNN\ = 2.76 TeV} \cite{Adam:2015rba, Khachatryan:2016ypw}. The error bars and open boxes around the data points represent statistical errors and systematic uncertainties for heavy-ion analyses, respectively. The boxes at unity from left to right show the global uncertainties for the LHC and STAR results. For this analysis, the global uncertainty includes the total uncertainties of the \pp\ reference. The theoretical calculations are for low and high \pT\ \jpsi\ at RHIC \cite{Yan:2006ve, Zhou:2014kka, Zhao:2010nk}. }
\label{fig:JpsiRaaVsCent}
\end{figure}

The dependence of the \jpsi\ suppression on collision centrality is shown in Fig. \ref{fig:JpsiRaaVsCent} as filled stars for $\pT>0.15$ \gev\ and $\pT>5$ \gev\ in \AuAu\ collisions at \mbox{\sqrtsNN\ = 200 GeV}.
It is measured in eight equally divided centrality intervals within 0-80\% at low \pT\ (left panel), while for the high \pT\ measurement the most peripheral bin is 60-80\% (right panel). The updated \RAA\ values from \jpsi's  measured through the dielectron channel are also shown as open circles in the right panel. At both low and high \pT, the \jpsi\ \RAA\ is seen to decrease from peripheral to central collisions, which is expected in the presence of the QGP. For high \pT\ \jpsi, where the CNM effects and the regeneration contribution are expected to be minimal, the \jpsi\ production in 0-10\% central collisions is suppressed by a factor of 3.1 with a significance of 8.5$\sigma$, providing strong evidence for the color-screening effect in the deconfined medium. Also shown in Fig. \ref{fig:JpsiRaaVsCent} is the \jpsi\ \RAA\ as a function of \npart\ measured for \PbPb\ collisions at \mbox{\sqrtsNN\ = 2.76 TeV} \cite{Adam:2015rba, Khachatryan:2016ypw}. Here, the low-\pT\ \jpsi's are above 0 \gev\ and high-\pT\ \jpsi's are above 6.5 \gev. Inclusion of very low-\pT\ \jpsi\ from coherent photoproduction in \PbPb\ collisions has negligible impact on the measured \RAA\ values for $\npart>50$ \cite{Adam:2015isa}. The low-\pT\ \jpsi's are much more suppressed in central and semi-central collisions at RHIC than at the LHC, likely due to the smaller charm quark production cross-section and thus smaller regeneration contribution at RHIC. On the other hand, the high-\pT\ \jpsi\ \RAA\ is systematically higher at RHIC for semi-central bins. This could be because the temperature of the medium created at the LHC is higher than that at RHIC, leading to a higher dissociation rate. Transport model calculations are consistent with the data at low \pT, while the data lay mostly between the two model calculations at high \pT\ except for the 60-80\% peripheral bin and the 0-10\% central collisions where the data point coincides with the TAMU model. The result from the Statistical Hadronization Model (SHM), shown as the dashed line in the left panel of Fig. \ref{fig:JpsiRaaVsCent}, also describes the data reasonably well in non-peripheral events \cite{Andronic:2017pug}. In the SHM model, the charm quark production cross-section from the fixed-order plus next-to-leading logs calculations \cite{Cacciari:1998it} is used as input. However, feed-down contributions from b-hadron decays have not been included.

\section{Summary}
In summary, we report the first measurements of inclusive \jpsi\ \RAA\ through the dimuon decay channel at mid-rapidity in \AuAu\ collisions at \mbox{\sqrtsNN\ = 200 GeV} by the STAR experiment at RHIC. Compared to previous dielectron measurements, the new results provide an improved measure of \jpsi\ suppression in the QGP with better precision and in a wider kinematic range. At low \pT, the interplay of the CNM effects, dissociation, and regeneration results in an increasing suppression of \jpsi\ from peripheral to central collisions. At \pT\ above 5 \gev, the \jpsi\ yield is significantly suppressed in central collisions, which is caused mainly by color-screening in the medium due to the presence of the QGP. While both the Tsinghua and TAMU transport models describe the centrality dependence of \jpsi\ \RAA\ at low \pT, their agreement with data degrades at high \pT. The new results presented in this letter will help constrain model calculations and deepen our understanding of the QGP properties. 

\section*{Acknowledgements}
We thank the RHIC Operations Group and RCF at BNL, the NERSC Center at LBNL, and the Open Science Grid consortium for providing resources and support.  This work was supported in part by the Office of Nuclear Physics within the U.S. DOE Office of Science, the U.S. National Science Foundation, the Ministry of Education and Science of the Russian Federation, National Natural Science Foundation of China, Chinese Academy of Science, the Ministry of Science and Technology of China and the Chinese Ministry of Education, the National Research Foundation of Korea, Czech Science Foundation and Ministry of Education, Youth and Sports of the Czech Republic, Hungarian National Research, Development and Innovation Office (FK-123824), New National Excellency Programme of the Hungarian Ministry of Human Capacities (UNKP-18-4), Department of Atomic Energy and Department of Science and Technology of the Government of India, the National Science Centre of Poland, the Ministry  of Science, Education and Sports of the Republic of Croatia, RosAtom of Russia and German Bundesministerium fur Bildung, Wissenschaft, Forschung and Technologie (BMBF) and the Helmholtz Association.

\section*{References}

\bibliography{mybibfile}

\end{document}